\documentclass[12pt]{iopart}
\usepackage{graphicx}

\begin{document}

\title{On the perturbations on satellites probing General Relativity}

\author{S.Sargsyan, G.Yegorian, S.Mirzoyan}

\address{Center for Cosmology and Astrophysics, 
Alikhanian National Laboratory, Yerevan, Armenia}
\ead{seda@yerphi.am}
\begin{abstract}
Non-gravitational Yarkovsky-Rubincam effect for LAGEOS and LAGEOS 2 satellites used to probe General Relativity has been
revealed by means of the Kolmogorov analysis of their perturbations. We present the method and its efficiency at modeling 
of generated systems with properties expected at the satellite laser ranging measurements and then at 
satellite residual data analysis.   

\end{abstract}

\maketitle

\section{Introduction}

Two Earth's satellites, LAser GEOdynamics Satellites (LAGEOS and LAGEOS2), have been used for testing of Lense-Thirring effect predicted by General Relativity, with resulting accuracy of 10\% \cite{CP,Ciufolini}. The recently launched satellite LARES (LAser RElativity Satellite) is aimed for even higher accuracy testing of Lense-Thirring effect \cite{Lares}. 

The analysis \cite{GC} of the LAGEOS and LAGEOS 2 residual data of their trajectories, i.e. the differences between real measurements and theoretically predicted trajectories based on the Earth's gravity field's possibly accurate reconstruction, has been performed using Kolmogorov method \cite{K,Arn,Arn1}. 
Kolmogorov analysis of the LAGEOS data revealed higher degree of randomness for LAGEOS than for LAGEOS 2 data, which was explained
via the Yarkovsky-Rubincam effect, i.e. thermal thrust due to the thermal radiation by the anisotropic thermal heating of a satellites by Solar and Earth's radiation. Although both satellites have almost identical orbits and internal structure, since LAGEOS has been 
on the orbit 16 years longer than LAGEOS 2, the effect of thermal thrust is random in the
case of LAGEOS with respect to LAGEOS 2. In fact, the thermal drag is a thermal acceleration of a satellite 
directed along the spin axis of a satellite. However, whereas the orientation of the LAGEOS spin axis was almost chaotic 
at the time of the orbital analysis,the spin of LAGEOS 2 had a more stable orientation over the same period, 
since LAGEOS 2 was launched much later than LAGEOS (16 years). This explains the more chaotic nature of 
the LAGEOS residuals.  

Below, we represent the method and its application to generated systems with properties expected for the LAGEOS residuals, which then enabled the analysis of the real data.

\section{The method vs the generated systems}

Kolmogorov's stochasticity parameter is defined for $n$ independent values $\{X_1,X_2,\dots,X_n\}$ of a variable $X$, given in increasing order \cite{K,Arn,Arn1}. Two
distribution functions are defined as follows: cumulative distribution function is 
$F(x) = P\{X\le x\}$, while $F_n(x)$ is the empirical distribution function 

\begin{equation}
F_n(x)= \left\{
\begin{array}{rl}
	0, & X < x_1 \\
	k / n, & x_k \leq X < x_{k+1} \\
	1, & x_n \leq X.\\
\end{array}
\right.
\label{eq:empiricdistribution}
\end{equation}

Then the stochasticity parameter $\lambda_n$ is 

\begin{equation}\label{KSP}
\lambda_n=\sqrt{n}\ \sup_x|F_n(x)-F(x)|\ .
\end{equation}
Kolmogorov's theorem states that for any continuous $F$ the limit $\lim_{n\to\infty}P\{\lambda_n\le\lambda\}=\Phi(\lambda)$
is converging uniformly and independent on $F$ and
\begin{equation}
\Phi(\lambda)=\sum_{k=-\infty}^{+\infty}\ (-1)^k\ e^{-2k^2\lambda^2}\ ,\ \  \lambda>0\ ,\label{Phi}
\end{equation}
where $\Phi(0)=0$.

For large enough $n$ and random sequence $x_n$ the stochasticity parameter $\lambda_n$ has a distribution tending to $\Phi(\lambda)$. If the sequence is not random, the distribution is different, therefore, the $\Phi(\lambda)$ defines the degree of randomness of a sequence \cite{Arn,Arn1}.

In order to show how this method can be informative for datasets as those of LARES satellites, we represent some results on the study of generated sequences revealing the behavior of $\lambda_n $ for certain classes of sequences. In Fig. 1 we represent the results for random and non-random sequences given as $x_n = 107 n\, mod513$ with randomly chosen coefficients; computations have been performed for 7,000 sequences of 10,000 elements each \cite{mod2}. 
The abscissa axes denote  $\lambda$ and the ordinate ones denote $\Phi(\lambda)$, the dashed line is the empirical distribution function, the solid line is the $\Phi(\lambda)$. From the Fig. 1a one can see that only for random sequences $\Phi(\lambda)$ practically coincide with the empirical distribution function, as follows from Kolmogorov's theorem. In Fig.1b  $\Phi(\lambda)$  is completely different from the empirical distribution function, as the values of $\lambda$ vary between 0.195 and 0.230.
 
\begin{figure}[ht]
\includegraphics[width=40pc]{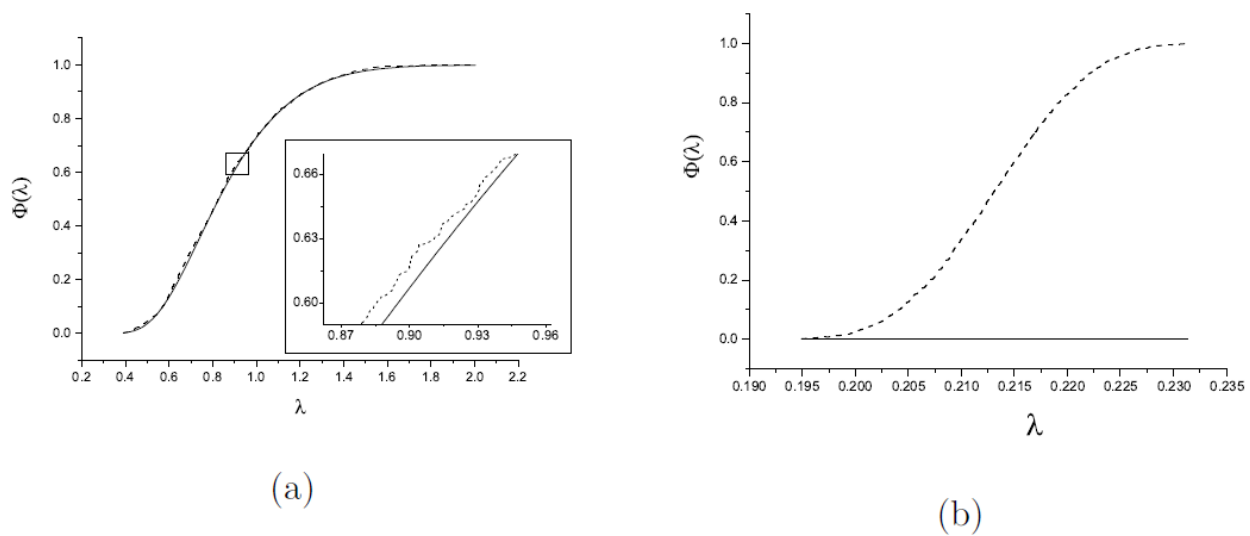}\hspace{1pc}%
\caption{The function $\Phi$ vs stochasticity parameter $\lambda$ for the system  $x_n = 107 n\, mod513$.}
\end{figure}

Another class of generated sequences included those given in the form $z_n = \alpha x_n + (1-\alpha) y_n$, so that  $x_n$ are the random sequences and $y_n = \frac{an\pmod b}{b}$ are the regular ones, $a$ and $b$ are prime numbers which are being fixed in each particular run, and both sequences within the interval $(0,1)$ have uniform distribution \cite{mod3}. Parameter $\alpha$ is representing the fractions of random and regular sequences, respectively. Fig. 2 shows the dependence of $\chi^2$ vs the parameter $\alpha$.

 \begin{figure}[ht]
\includegraphics[width=20pc]{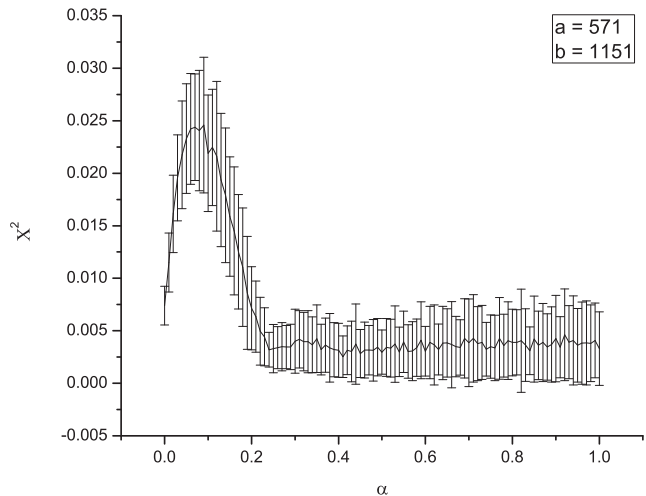}\hspace{2pc}%
\caption{The behavior of $\chi^2$ vs the parameter $\alpha$ denoting the fraction of random and regular subsequences.}
\end{figure}

These examples illustrate how the method works for sequences with various degree of correlations and randomness. This method has been applied for the analysis of the properties of cosmic microwave background radiation, data on X-ray clusters (\cite{G2009,G} and references therein).

\section{The satellite residuals}

The data for LAGEOS and LAGEOS 2 which carried sets of laser reflectors, have been collected via the laser ranging from ground based stations during about 11 years, i.e. 4018 days by step of 14 days {\cite{CP,Ciufolini}. The residuals have been obtained using the measured data and the theoretical trajectories obtained using the modeled gravity field of the Earth (Fig.3). 

\begin{figure}[ht]
\includegraphics[width=20pc]{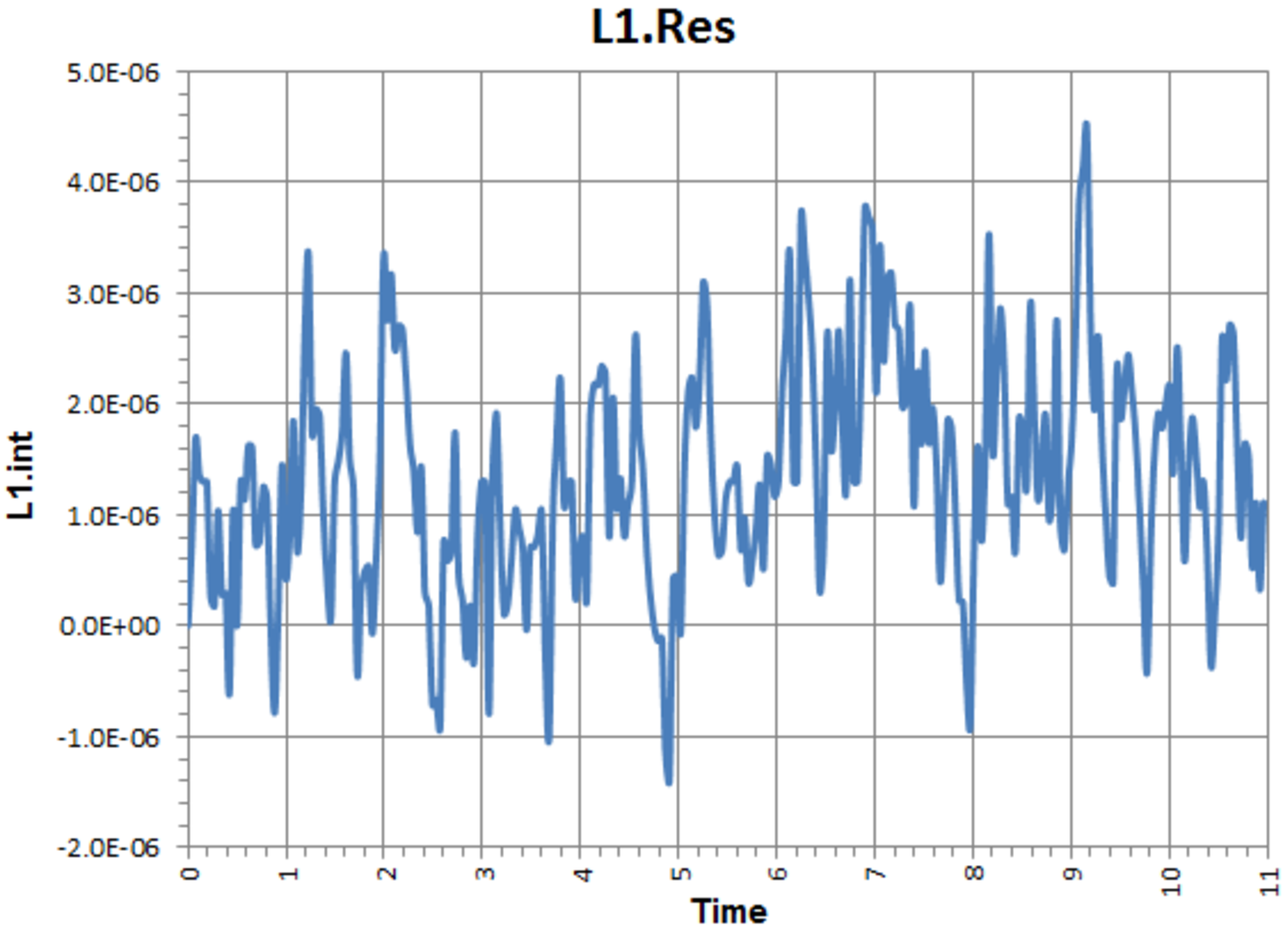}\hspace{2pc}%
\includegraphics[width=20pc]{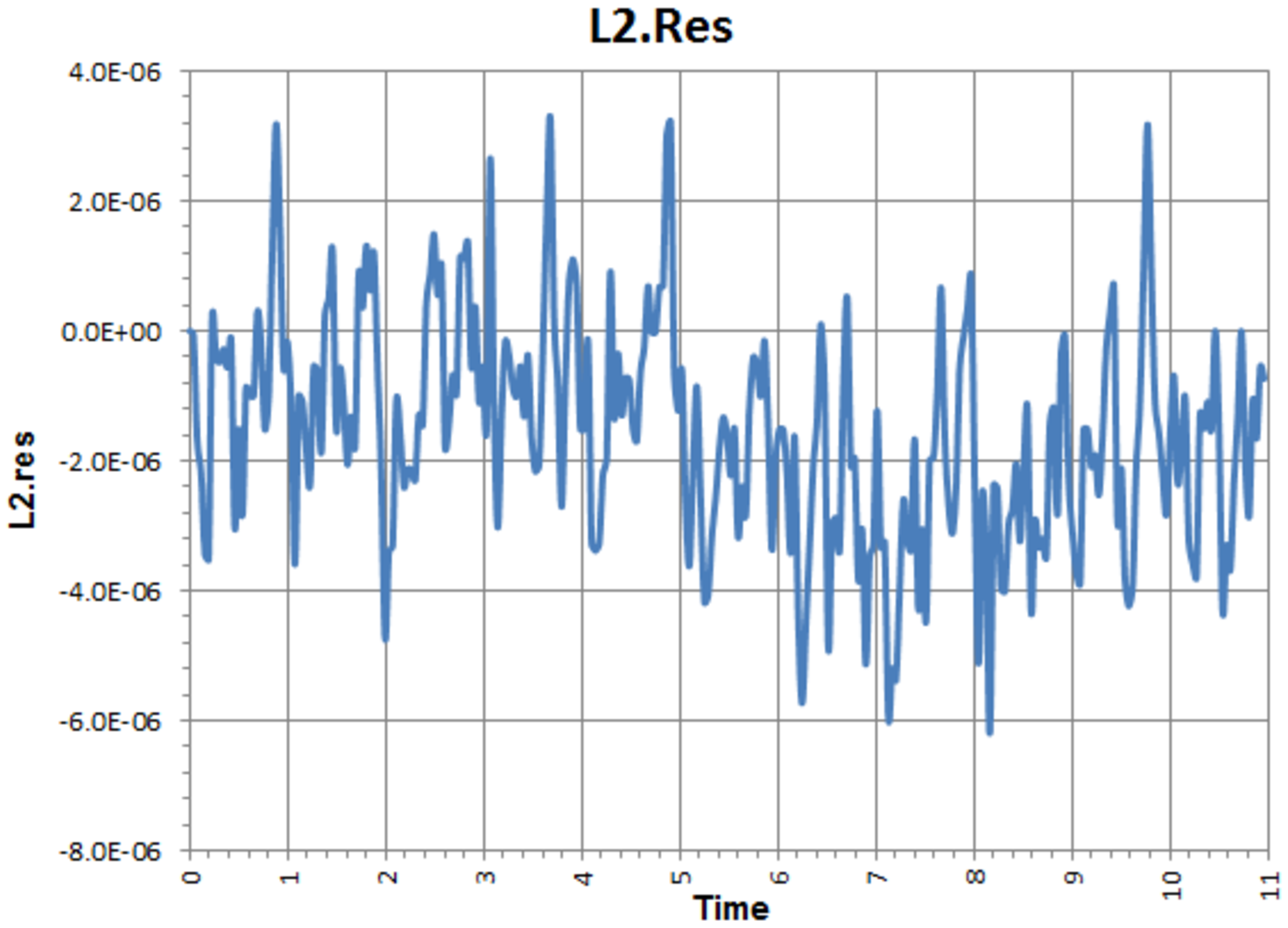}\hspace{2pc}%
\caption{\label{fig:res} The residuals of LAGEOS and LAGEOS 2.}
\end{figure}

Both satellites have almost identical semi-major axes - for LAGEOS it equals to 12270 km and for LAGEOS 2 to 12160 km - with different orbital inclinations. The important difference is their stay time on the Earth's orbit; LAGEOS and LAGEOS2 have been launched on 4 May 1976 and 23 October 1992, correspondingly.

We used Kolmogorov's method to analyse the residual datasets of both LARES satellites, to obtain their comparative degree of randomness.
Despite the fact that both satellites are identical, one finds non-identical behavior of $\Phi$ for LAGEOS and LAGEOS 2 for Gaussian CDF vs the variation of the standard deviation $d\sigma$ {\cite{GC}} (Fig.\ref{fig:lg}). As shows Fig. 4 the residuals of LAGEOS do possess about 10 times higher degree of randomness (chaos) than of LAGEOS 2.

Thus, the behavior of  $\Phi$ function for two identical satellite residuals enables to reveal Earth-Yarkovsky or Yarkovsky-Rubincam effect \cite {R}, i.e. non-gravitational perturbations acting on the satellites. The possibility of accurate estimation of the contribution of such non-gravitational effects is crucial for the basic goal, i.e. highly accurate testing of predictions of the General Relativity and obtaining constraints on its extensions.

\begin{figure}[ht]
\includegraphics[width=19pc]{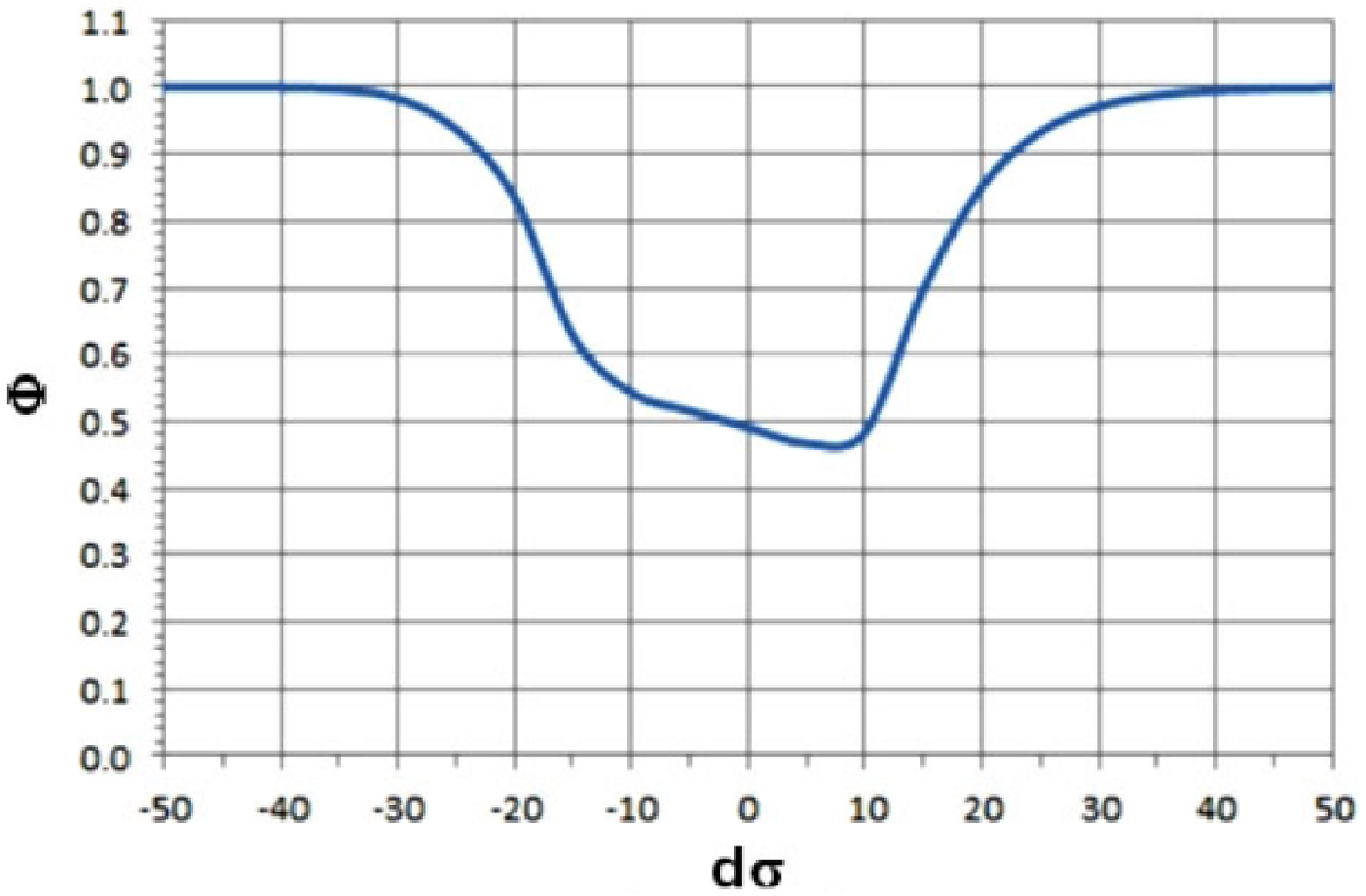}\hspace{2pc}%
\includegraphics[width=19pc]{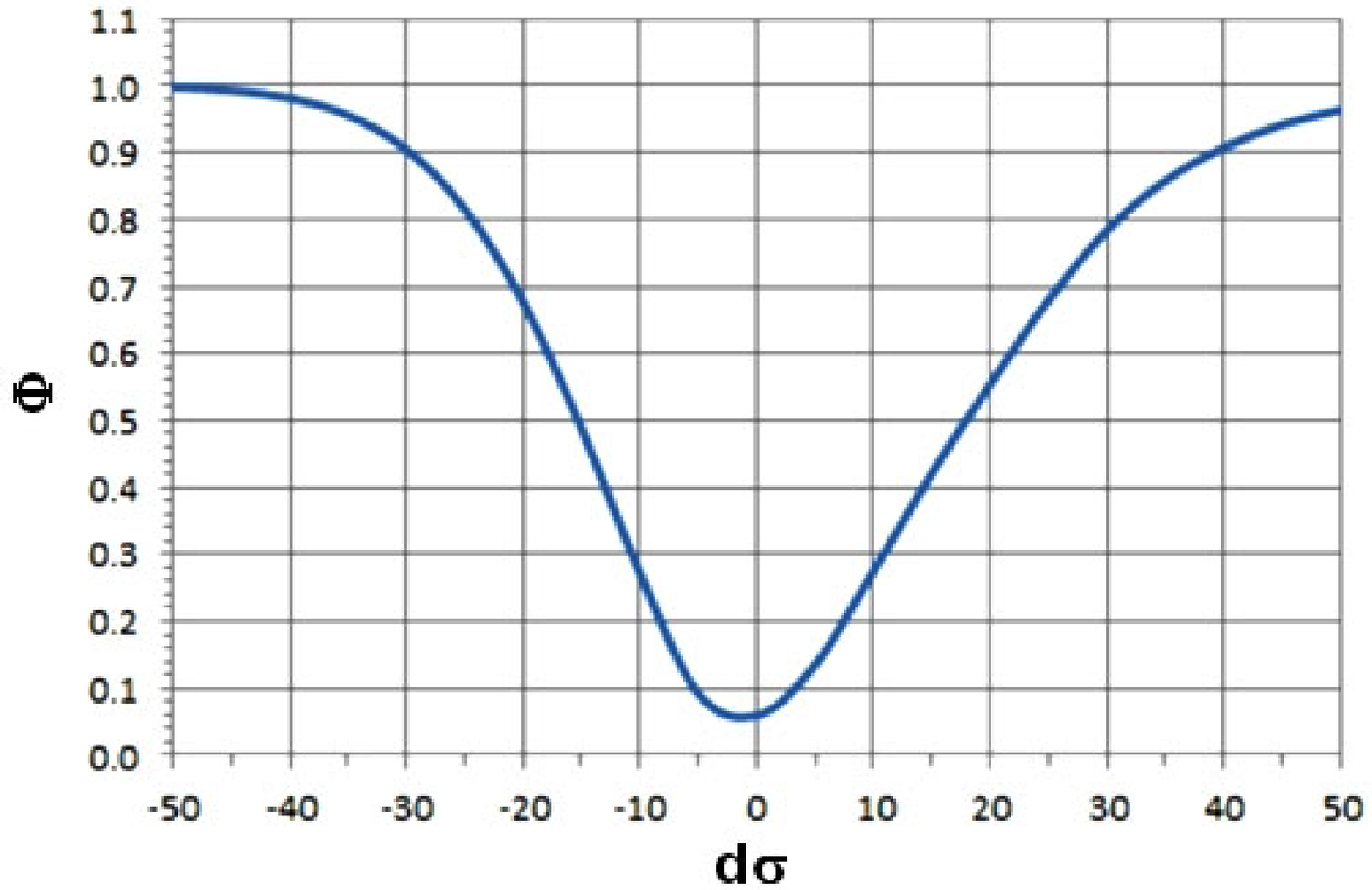}\hspace{2pc}%
\caption{\label{fig:lg} Kolmogorov function $\Phi$ vs $d\sigma$, LAGEOS (left), LAGEOS 2 (right).}
\end{figure}

We thank I.Ciufolini, V.Gurzadyan, A.Paolozzi for the joint work on LAGEOS data. S.S. thanks G.Meylan for the hospitality in Laboratoire d'Astrophysique, EPFL, Lausanne, during the work on this paper.

\end{document}